\documentclass[sigplan,10pt,nonacm,pbalance]{acmart}

\setcopyright{none}
\renewcommand\footnotetextcopyrightpermission[1]{}
\usepackage{tikz}
\usepackage{amsmath}
\usepackage{caption}

\newif\ifsubmission
\submissiontrue
\newcommand{\mcnote}[1]{}

\begin{document}

\date{}

\title[Cache-Resident LLM Inference in GB-Scale Last-Level Caches]{%
    Cache-Resident LLM Inference \texorpdfstring{\\}{} in GB-Scale
    Last-Level Caches
}

\author{Wanning Zhang}
\affiliation{%
    \institution{King Abdullah University of Science and Technology}
    \country{Saudi Arabia}
}
\email{wanning.zhang@kaust.edu.sa}

\author{Tongzhou Gu}
\affiliation{%
    \institution{King Abdullah University of Science and Technology}
    \country{Saudi Arabia}
}
\email{tongzhou.gu@kaust.edu.sa}

\author{Marco Canini}
\affiliation{%
    \institution{King Abdullah University of Science and Technology}
    \country{Saudi Arabia}
}
\email{marco@kaust.edu.sa}

\author{Ceyu Xu}
\affiliation{%
    \institution{The Hong Kong University of Science and Technology}
    \country{Hong Kong SAR, China}
}
\email{eeentropy@ust.hk}

\author{Jian Weng}
\affiliation{%
    \institution{King Abdullah University of Science and Technology}
    \country{Saudi Arabia}
}
\email{jian.weng@kaust.edu.sa}

\renewcommand{\shortauthors}{Zhang et al.}


\begin{abstract}
\emergencystretch=2em
Large language model (LLM) inference is increasingly dominated by data
movement across the memory hierarchy.
Recent advances in 3D-stacked cache technologies have enabled GB-scale
last-level caches in modern server CPUs.
These caches make it possible to keep reusable model weights on chip
and exploit cache bandwidth and latency.
Achieving this regime, however, is not straightforward.
Under conventional pipelining, the deeper partitioning needed for weight
residency also increases in-flight requests and therefore the KV-cache
footprint.
Meanwhile, once operators become cache-resident, operator-centric
execution pays disproportionate synchronization overhead at every
operator boundary.

We present a cache-resident execution model for inference on
hierarchical-memory clustered systems.
The model separates weight-centric operators from attention and KV-cache
management into dedicated resource domains.
This keeps reusable weights cache-resident while scaling KV capacity
independently of pipeline depth.
It also relaxes synchronization from operator boundaries to true
sub-operator dependencies, reducing coordination overhead in the
cache-resident regime.

We instantiate this model on a multi-socket CPU cluster with a
weight--attention decoupled architecture, locality-aware placement,
and a specialized static runtime.
Our evaluation shows that the prototype substantially outperforms equally
provisioned \texttt{llama.cpp}.
Measured end to end on the deployed Llama-3.2-3B and Llama-2-7B
configurations, it achieves $2.04\times$--$11.51\times$ speedup on
time-per-output-token (TPOT).
Under a validated analytical model, it further reaches up to
$13.9\times$ TPOT speedup across a broad range of model sizes, context
lengths, and batch sizes.
These results show that commodity CPUs with GB-scale last-level caches
can support efficient LLM inference when execution is organized around
cache residency, decoupled state management, and dependency-aware
coordination.

\end{abstract}

\maketitle

\section{Introduction}
\label{sec:intro}

Large language model (LLM~\cite{attention}) inference is prominently
dominated by data movement.
Each token is decoded by propagating embeddings through operators layer
by layer, involving activated model weights and attention state
(e.g., KV cache).
As model sizes and context lengths grow, the repeated decoding incurs
substantial data movement across the memory hierarchy and frequent
synchronization at operator boundaries,
leading to compute unit underutilization.

A natural, if ideal, response is to keep model weights resident in
on-chip caches.
To do so, weights must be partitioned across multiple cache instances.
Layer-wise partitioning, a.k.a. pipeline parallelism (PP), shards the
model by layers, so the pipeline depth is determined by the model size
relative to cache capacity.
Although a deeper pipeline reduces per-stage weight footprint,
it also increases the number of concurrently in-flight tokens.
Attention state (KV cache) must be maintained for every in-flight token at each stage.
As pipeline depth increases, the total KV cache footprint grows proportionally,
competing for the same limited cache capacity required for weight residency.
Since both attention state and model weights contend for the same cache resources,
a fundamental scalability limitation emerges for cache-resident deployment under PP.

Our insight is that weights and attention have fundamentally
different access and reuse patterns.
Weights are static and can be reused for different tokens, while the
KV cache is a runtime state dedicated to a single query.
This cache-resource contention can be eliminated by re-organizing
weights and attention state into separate dedicated resources.

In contrast, intra-layer partitioning, a.k.a. tensor parallelism (TP),
splits individual operators within a layer across different computing domains
to reduce memory pressure and improve utilization.
In practice, TP is implemented under an operator-centric execution model:
the data-parallel loops of each operator are distributed across domains,
followed by a synchronization barrier to materialize results
before subsequent operators proceed.
Although convenient for modular system design, such global
synchronization becomes increasingly costly when data already reside in
on-chip cache and computation is fine-grained~\cite{tofu}.

Our insight is that this synchronization overhead can be substantially reduced
by adopting a finer-grain scheduling strategy.
Many sub-computations within a layer (e.g. attention heads)
are semantically independent and do not require immediate global synchronization.
Scheduling computation at sub-operator granularity avoids
unnecessary barriers and amortizes synchronization overhead.

These ideas are realized in a cache-resident execution model for
cache-resident inference on hierarchical-memory clustered systems.
The model is prototyped on a multi-socket CPU architecture with
large shared last-level caches.
High-end server-class multi-socket CPUs already provide sufficient
arithmetic throughput for memory-dominated LLM inference.
Once data resides in on-chip cache,
general-purpose hardware can achieve substantial efficiency gains.

The technical contributions of this work are as follows:

\begin{itemize}
    \item \textbf{Performance bottleneck characterization under cache-residency.}
    Inference is typically bounded by data movement across
    the memory hierarchy that serves the working set, and
    co-located weights and KV state amplify this cache pressure.
    Even when cache residency is achieved,
    the synchronization overhead at
    operator boundaries becomes a bottleneck.

    \item \textbf{An execution model for scalable cache-resident inference.}
    The model decouples weight-centric and attention-centric operators
    to optimize each individual access and reuse pattern, and
    relaxes synchronization from operator boundaries to true
    data dependencies.

    \item \textbf{A practical implementation and evaluation.}
    An instantiation on a multi-socket CPU cluster validates this
    execution model, demonstrates substantial performance gains over
    \texttt{llama.cpp}, and exposes the fixed-resource tradeoff of
    WA separation.
\end{itemize}

The execution bottlenecks and resource-organization principles developed
in this work are expressed at the level of memory hierarchy and
execution dependencies, and are therefore not specific to CPUs.
We instantiate them on CPUs for two reasons.
First, GB-scale last-level caches make modern multi-socket CPUs a
natural substrate for studying cache-resident inference.
Second, CPUs remain broadly available deployment hardware even when
high-end accelerators, e.g. GPUs, are scarce because of supply chain capability
and geopolitical regulations~\cite{bis-export}.

Our evaluation shows that the prototype substantially outperforms
equally provisioned \texttt{llama.cpp}.
Measured end to end on the Llama-3.2-3B and Llama-2-7B deployments,
it achieves $2.04\times$--$11.51\times$ TPOT speedup.
Under model-based extrapolation, it further reaches up to
$12.5\times$ throughput speedup and up to $13.9\times$ TPOT speedup
across model sizes, context lengths, and batch sizes of Llama-family
models.

\section{Background \& Motivation: LLM Inference}
\label{sec:bg}

In this section, we first characterize transformer-based LLM inference.
We then discuss a hardware-agnostic paradigm for clustered inference
systems by analyzing its performance model, scalability limits, and the
programming abstraction that motivates this work.

\begin{figure}
    \centering
    \includegraphics[width=0.85\linewidth]{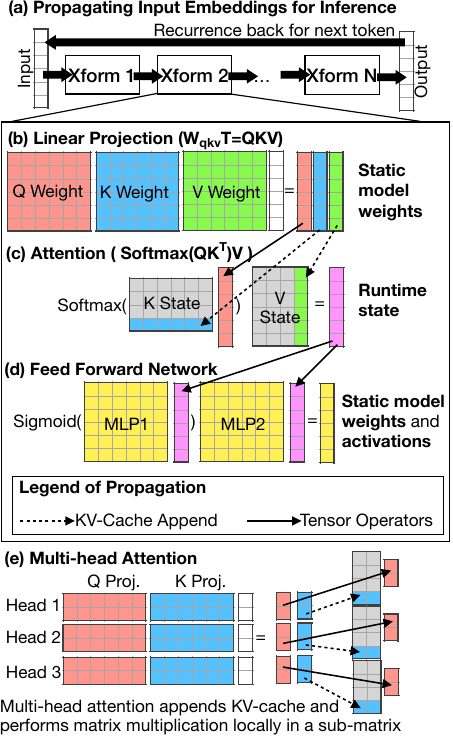}
    \Description{Transformer block diagram showing QKV projection, attention, and FFN components.}
    \vspace{-0.1in}
    \caption{Transformer-based LLM decodes each token by propagating the embeddings across operators.}
    \label{fig:bg-xform}
    \vspace{-0.1in}
\end{figure}

\subsection{Transformer-based LLM Inference}

Fig.~\ref{fig:bg-xform}(a) illustrates a transformer-based LLM inference:
When decoding, each token is generated by propagating an embedding vector
through a stack of transformer blocks.
The output embedding of the final block is then projected to produce the next token,
which is fed back into the model for subsequent steps.

Our system supports both prefilling and decoding
(see \S\ref{sec:eval}), but the rest of the paper focuses on
decoding because it is the long-running steady state and dominates
execution time.

Fig.~\ref{fig:bg-xform}(b)--(d) zooms into a single transformer block,
which typically consists of three phases: QKV projection, attention,
and feed-forward network (FFN).

\begin{figure}[t]
    \centering
    \includegraphics[width=0.65\linewidth]{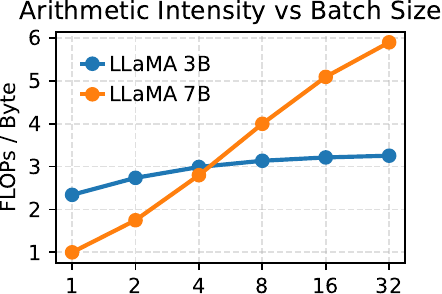}
    \Description{Bar chart showing FLOPs/Byte ratio for Llama-2 3B and 7B models.}
    \vspace{-0.15in}
    \caption{
    FLOPs/Byte of Llama-2 3B \& 7B during decoding
    with context length $4096$.
    }
    \label{fig:arithmetic-intensity}
    \vspace{-0.15in}
\end{figure}

\noindent\textbf{Common Linear Operations.}
The dominant computations across all transformer inference
phases are matrix--vector multiplications
(or thin matrix--matrix multiplications when batched).
Such kernels provide very limited data reuse because each
weight parameter is typically consumed only once per token.
Batching is commonly applied to mitigate such scarce data reuse,
allowing shared model weights to be reused across multiple
requests. However, sequence-specific runtime states (i.e.,
the KV-cache) must still be accessed independently for each
sequence. Consequently, as batch size increases, the effective
arithmetic intensity improves only modestly and remains relatively
low, as shown in Fig.~\ref{fig:arithmetic-intensity}.
This observation suggests that decoding performance is still
largely constrained by memory bandwidth, leaving substantial
opportunities for improving data locality.

\noindent\textbf{Static-weight operators.}
The QKV projection and FFN depend solely on model weights,
which remain constant across different queries.
Therefore, the data movement can be amortized by batching queries,
which increases arithmetic intensity and reuses weights.
However, batch size cannot be arbitrarily large because of the
latency requirements and cache footprint for operator optimization.

\noindent\textbf{State-dependent attention.}
In contrast, attention introduces a dependence on dynamically
growing runtime state, namely the KV cache~\cite{one-write-head,pagedattention}.
For each token, the query vector interacts with the accumulated KV states
in the cache. Prior KV-vectors were stored in this cache,
and newly projected KV vectors are appended to this cache.
Unlike projection or FFN operators, whose weights are
shared across queries, attention involves per-sequence state
that grows with context length.
Consequently, batching different queries
does not increase data reuse within attention.

\emph{Takeaway. Decoding a single token under a conventional
transformer-based LLM accesses all activated model weights
along the forward path as well as the related accumulated KV
cache for attention computation. This process is prominently
dominated by data movement due to the low data reuse.}

\begin{figure}
    \centering
    \includegraphics[width=1.05\linewidth]{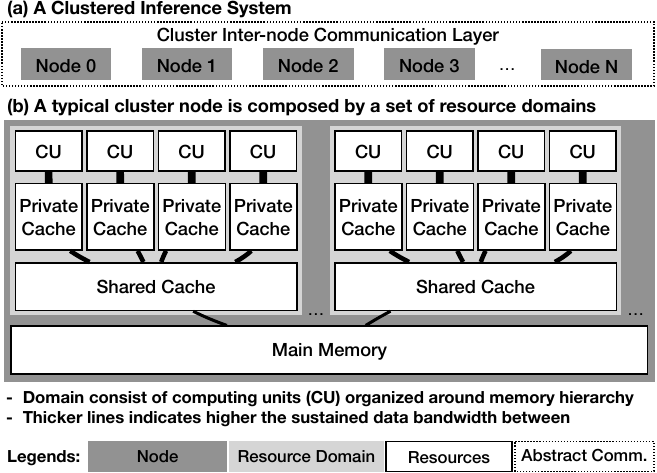}
    \Description{System architecture diagram showing hierarchical memory organization across nodes, CCDs, and cores.}
    \vspace{-0.3in}
    \caption{Architectural paradigm for LLM inference system.}
    \label{fig:bg-sys}
    \vspace{-0.2in}
\end{figure}

\subsection{A Paradigm of Clustered Processing Systems}\label{sec:sys-paradigm}

Fig.~\ref{fig:bg-sys} presents a high-level architectural paradigm
commonly used for large-scale LLM inference.
The abstraction is intentionally hardware-agnostic, without binding to
a specific platform or topology.

Computational units (CU) are organized around a hierarchical memory
system to sustain data for processing.
The hierarchy consists of private caches and access to a shared
last-level cache (LLC), backed by main memory.
Higher levels in the memory hierarchy typically provide higher
bandwidth but lower capacity.
This bandwidth-capacity imbalance requires careful control of the
active working set to maximize data reuse and CU utilization.

A group of computing units sharing portions of the memory hierarchy
forms a \emph{resource domain}.
Within a resource domain, communication bandwidth is significantly
higher than across domains, as illustrated by the thicker
interconnect links in Fig.~\ref{fig:bg-sys}(b).

Multiple resource domains are integrated into a \emph{node}.
In practice, node size is bounded by physical design constraints,
including chip area, packaging, power, and manufacturing yield.
Thus, a single node cannot scale arbitrarily in compute
or cache capacity.

To scale beyond the limits of a single node, multiple nodes are
connected via an inter-node communication layer, forming a cluster.
Inter-node bandwidth is typically lower and higher latency than
intra-node communication, introducing additional costs for
synchronization and data movement.

This hierarchical organization introduces fundamental tradeoffs among
computation placement, data reuse, and communication cost, which we
leverage throughout the rest of the paper.

Further, this scaling abstraction is intentionally separated from any
specific hardware instantiation.
However, in this work we realize and evaluate it on a multi-socket CPU
cluster, whose GB-scale last-level caches make it a natural substrate
for studying cache-resident LLM inference.

\begin{figure}
    \centering
    \includegraphics[width=0.99\linewidth]{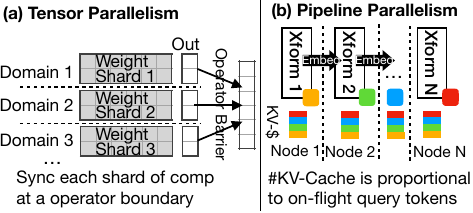}
    \Description{Illustration of tensor parallelism and pipeline parallelism strategies.}
    \vspace{-0.15in}
    \caption{Tensor and pipeline parallelism (TP \& PP) scaling}
    \label{fig:parallelism}
    \vspace{-0.15in}
\end{figure}

\subsection{Performance Scaling} \label{sec:scalability}

We analyze performance scaling~\cite{tofu,scaling-inference} on the
clustered system above using two widely adopted techniques,
tensor parallelism (TP) and pipeline parallelism (PP), as illustrated
in Fig.~\ref{fig:parallelism}. TP shards individual operators across
devices~\cite{megatron-lm}, while PP partitions consecutive layers into
pipeline stages~\cite{gpipe,pipedream}.
Inference performance scales along two coupled dimensions:
\textit{throughput} and \textit{latency}.

\noindent\textbf{Parallelization Strategy.}
Fig.~\ref{fig:parallelism}(a) illustrates tensor parallelism,
which distributes independent output computations of an operator
across multiple resource domains,
each has its own CUs and memory hierarchy.
Partial results are synchronized at operator boundaries.

In this work, we do not apply TP across nodes,
as cross-node synchronization incurs substantial communication overhead.
Instead, pipeline parallelism (PP) is used to scale across nodes
by assigning consecutive transformer layers to different nodes.
Only embeddings are exchanged between nodes,
reducing inter-node communication volume.

\noindent\textbf{Performance Upper Bound.}
We build an analytical model to characterize latency and throughput
under PP scaling.
Transformer-based LLMs are constructed by stacking nearly identical layers,
allowing layers to be evenly partitioned across pipeline stages (nodes).
Let the latency of a single evenly partitioned pipeline stage be $l$,
and let the pipeline depth be $p$.
The end-to-end latency of decoding one token is:
$L = p \cdot l$. Under steady-state execution, throughput is bounded by the slowest stage: $T = \frac{1}{l}$.

Since transformer matrix-vector multiplication exhibits
low arithmetic intensity and limited data reuse,
the service time of a pipeline stage is typically memory-bound.
Thus,
\[
l \ge \frac{\text{Memory Footprint per Token}}{\text{Effective Memory Bandwidth}},
\]
where memory footprint per token
includes model weights and runtime state,
and the effective bandwidth of the highest memory level
that accommodates this footprint.

\noindent\textit{\textbf{Opportunity 1: Cache-Resident Inference.}}
A straightforward response to the model above is to place
the active working set in the highest-bandwidth memory level available.
Recent advancements in 3D-stacked cache technologies
have enabled GB-scale on-chip caches.
Accommodating both model weights and runtime state on chip
could significantly improve the per-stage serve time.
However, achieving cache residency introduces a fundamental
paradox of scalability.

\noindent\textit{\textbf{Challenge 1:  KV-Cache Pressure Paradox.}}
Cache-resident inference becomes fundamentally difficult when scaling
to \emph{larger models}. Specifically, KV-cache is request-specific,
so each KV-cache is separately maintained, and the total KV-cache scales with:
$$ \text{KV Cache} \propto \#\text{Layers}\times\text{Batch}\times\text{(Ctx Len.)} $$

Scaling the model size increases the number of transformer layers,
and the associated KV-cache sizes.
Meanwhile, the model weights should also be partitioned across more pipeline stages, $p$.
To sustain steady-state throughput, at least $p$
decoding requests must remain in flight to keep all stages busy.
Assuming each node is evenly responsible for $\frac{\#\text{Layers}}{p}$ layers,
the per-node KV-cache pressure is:

\[
\begin{array}{rcl}
\text{Per-node KV Cache}
&\propto& \dfrac{\#\text{Layers}}{p}\times p\times \text{Batch}\times \text{(Ctx Len.)} \\
&\propto& \#\text{Layers}\times \text{Batch}\times \text{(Ctx Len.)}
\end{array}
\]

\emph{Takeaway. A paradox emerges: assigning fewer layers to each
pipeline node reduces the local weight footprint, but it also increases
the number of pipeline stages and therefore the number of in-flight
requests needed to keep the pipeline busy. These effects cancel, so the
per-node KV-cache pressure remains unchanged. Simply allocating layers
across more pipeline stages does not relieve cache pressure under a
weight--KV co-located design.}

This observation also guides our evaluation methodology.
End-to-end evaluation on real machines is preferable, but limited
hardware makes it practical to profile isolated per-node experiments
under matched KV-cache pressure and then estimate end-to-end
performance with an analytical model.
See \S\ref{sec:eval} for details.

\noindent\textit{\textbf{Challenge 2: Synchronization Overhead.}}
When execution is memory-bound, synchronization overhead is dwarfed by
memory latency. Once cache residency is achieved, operator kernels
become much shorter, and synchronization is no longer negligible.
All machine learning frameworks~\cite{tvm,tensorflow,pytorch} adopt an
operator-centric model.
Operators serve both as units for model development and as boundaries
for synchronization and scheduling.
All parallelized shards must synchronize before downstream operators can
proceed.

\subsection{Operator-centric Execution} \label{sec:operator-centric}

The operator-centric abstraction has been widely adopted in machine
learning frameworks~\cite{tvm,tensorflow,pytorch,halide} because it
offers a near one-to-one correspondence between mathematical
formulations and execution units.
This abstraction allows model developers and system engineers to focus
on their respective scopes, but it also imposes an abstraction tax.
Each operator forms a strict synchronization boundary:
all partial results must be fully synchronized
before feeding to downstream operators begin.
While operator fusion mitigates overhead for simple elementwise functions 
(e.g., sigmoid or exponential), it is ineffective for structurally 
independent matrix multiplications within transformer blocks.

\noindent\textbf{\textit{Opportunity 2: Head Independence.}}
In LLM inference, many computations are semantically independent
at the attention-head granularity.
As illustrated in Fig.~\ref{fig:bg-xform}(e),
each head operates on its own shard of QKV projections
and corresponding KV state, and can compute attention independently.
Nevertheless, operator-centric execution enforces
global synchronization at operator boundaries,
introducing redundant coordination beyond true data dependencies.

As operator kernels become smaller under cache-resident execution,
their latency decreases, and the relative cost of
cross-head coordination becomes increasingly significant.
This mismatch between true data dependencies and
operator-level synchronization limits scalability.

\section{Architecture} \label{sec:arch}

The discussion above already identifies both the scalability paradox
and the resulting synchronization overhead in cache-resident inference.
Our goal is to sustain cache residency of the entire active working set,
including reusable model weights and dynamically growing KV state,
within high-bandwidth on-chip memory.

We first introduce a weight--attention (WA) decoupled organization,
derived from the different behaviors of static weights and the growing
KV cache, and then redesign synchronization around sub-operator data
dependencies.

\begin{figure*}
    \centering
    \includegraphics[width=0.95\linewidth]{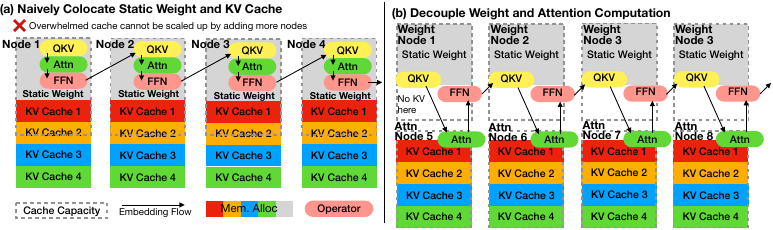}
    \Description{Architecture diagram showing weight-attention decoupling with separate weight nodes and attention nodes.}
    \vspace{-0.1in}
    \caption{Comparison between weight--KV colocation and
    weight--attention decoupling}
    \label{fig:wa-decouple}
    \vspace{-0.1in}

\end{figure*}

\subsection{Weight–Attention Decoupled Organization} \label{sec:wa-decouple}

As discussed in \S\ref{sec:scalability},
scaling to larger models requires deeper pipeline
to shard model weights across nodes.
However, deeper pipeline also enlarges the KV footprint,
since at least $p$ in-flight decoding requests are required
to sustain a pipeline of depth $p$.
As illustrated in Fig.~\ref{fig:wa-decouple}(a),
the growing KV caches and static weights
compete for the same cache capacity,
evicting reusable weight data to lower-level memory hierarchy with lower bandwidth.
Under such naive resource allocation, increasing the number
of nodes further deepens the pipeline
and proportionally amplifies the KV footprint.

Our insight into this issue lies in a key behavior
difference between weight-centric
and attention operators.
Weight-centric operators themselves (QKV projection and FFN) repeatedly
access reusable model weights
but do not depend on dynamically growing KV state.
In contrast, the attention operators,
expressed as $Attn(Q)=\mathrm{Softmax}(QK^\top)V$,
form a closed set that interacts exclusively with KV cache
and local activations, without accessing static weights.

Such difference reveals a natural boundary between
memory and compute resource allocation, as shown in
Fig.~\ref{fig:wa-decouple}(b).
Weight-centric operators are grouped into weight nodes that prioritize
reusable weight residency, and no KV cache is maintained there.
Attention-centric operators are instead isolated into attention domains
dedicated to managing KV state.
Additional KV-cache capacity can then be integrated by routing the
outputs of weight-centric operators to decoupled attention nodes,
enabling scaling to deeper pipelines or longer contexts.

\noindent\textbf{Scalability:} Because weight-centric and attention
operators are separated into different resource domains, KV-cache
management becomes independent of weight execution.
In our design, each decoding sequence is associated with a specific attention
node that owns its KV-cache state, and the outputs of weight-centric
operators are routed to that node.
This routing mechanism lets the system scale KV capacity by attaching
additional attention nodes without increasing the pipeline depth of the
weight-centric stages.
WA separation is optional: when KV-cache pressure is still modest, a colocated
design remains more socket-efficient; when latency is the priority, dedicating
an attention node removes KV interference from the weight-resident path.
Even if the KV-cache working set eventually exceeds L3 capacity, the resulting
performance degradation is confined to the corresponding attention node, while
weight-centric operators remain cache-resident on weight nodes and continue to
benefit from high-bandwidth cache access.
See our evaluation in \S\ref{sec:eval} for more details on the effects
of WA separation and sensitivity to cache overwhelming.

By accommodating the working set in high-bandwidth cache,
the compute pipeline stall caused by sustaining data is substantially
reduced. However, the synchronization overhead does
not shrink accordingly.
As discussed in \S\ref{sec:operator-centric},
operator-centric execution enforces dependencies at operator
granularity even though only a subset of them is required by the
underlying semantics.
This motivates rethinking synchronization granularity
beyond operator boundaries.

\subsection{Sub-operator (A)synchronicity}

\begin{figure}
    \centering
    \includegraphics[width=\linewidth]{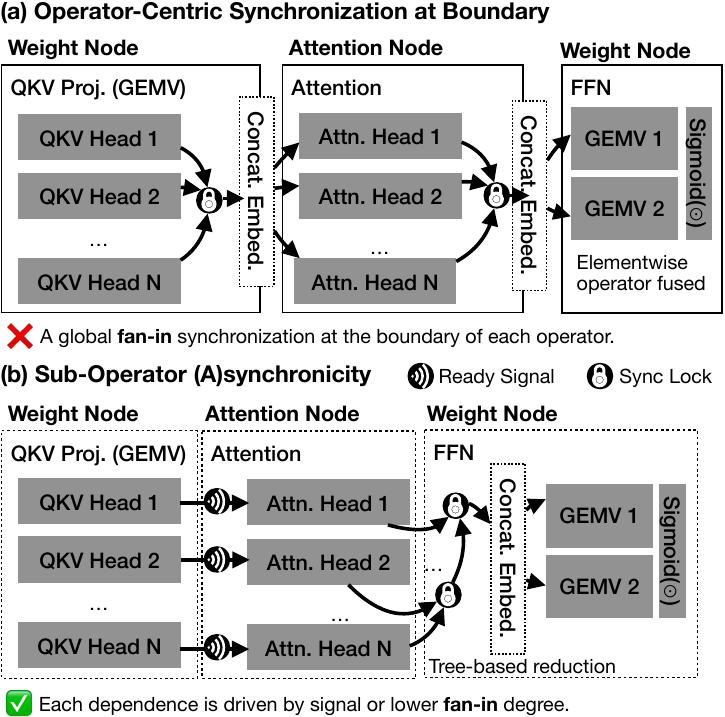}
    \Description{Diagram illustrating sub-operator dependency tracking and hierarchical synchronization.}
    \vspace{-0.25in}
    \caption{Sub-operator coordination}
    \label{fig:lock}
    \vspace{-0.2in}
\end{figure}

After cache residency is achieved, synchronization overhead
is no longer negligible. Under the operator-centric execution
model, each operator forms a strict synchronization boundary:
all tensor-parallel partial results must be aggregated
before proceeding. This enforcement creates
a global fan-in at every operator boundary, as shown in
Fig.~\ref{fig:lock}(a).
Such synchronization cost scales with fan-in: the more
tensor-parallel shards converge on the same completion state, the
higher the coordination overhead and contention on the shared lock
state.

However, the true data dependencies exist at sub-operator
granularity. In multi-head attention, each head operates
on its own shard of projections and KV state. One head
does not semantically depend on the completion of others.
Instead of synchronizing at the operator boundary,
we allow each head to propagate its readiness independently.

As illustrated in Fig.~\ref{fig:lock}(b), each head emits a ready
signal upon completion. Downstream computation is driven by these
signals rather than by a shared global barrier.

On the other hand, gated FFN,
$\sigma(M_1 v)\odot \sigma(M_2 v)$, requires the full embedding vector $v$.
This yields two options for coordinating partial progress across heads.
Each has tradeoffs between memory traffic and synchronization.
One option is to compute an outer-product-style partial update per
segment and then perform a single global synchronization to aggregate
each partial sum of $v$; however, this approach incurs large memory
traffic between L1 and LLC.
Instead, we perform a tree-based synchronization that bounds fan-in at
each merge and then execute fused GEMV and elementwise operations, as
shown in Fig.~\ref{fig:lock}(b).
See the next section for implementation details.

To sum up, synchronization granularity is aligned with true
data dependencies rather than operator abstraction boundaries.
Execution progresses according to per-head readiness and
bounded-degree accumulation, reducing coordination overhead
without altering semantic correctness.

\section{Implementation}

\begin{figure}
    \centering
    \includegraphics[width=0.95\linewidth]{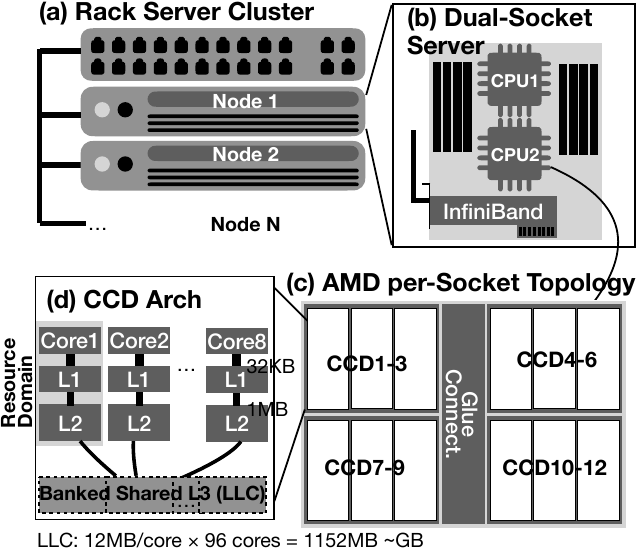}
    \Description{Hardware implementation showing dual-socket server with InfiniBand interconnection.}
    \vspace{-0.1in}
    \caption{Prototyping Hardware Platform}
    \label{fig:hw-platform}
    \vspace{-0.2in}
\end{figure}

This section describes how the proposed design is realized on the clustered system of
Fig.~\ref{fig:bg-sys} and the hardware platform of
Fig.~\ref{fig:hw-platform}.
We prototype it on a server cluster built from AMD EPYC 9684X CPUs.
Each server node contains two CPU sockets and connects to a central
InfiniBand switch through PCIe NICs.

\subsection{Resource Mapping for Parallelization}
\label{sec:resource-mapping}

\noindent\textbf{Pipeline Parallelism across Rack Nodes.}
At the cluster level, the system applies pipeline parallelism (PP)
by partitioning consecutive transformer layers across rack nodes.
Each rack node hosts multiple pipeline stages.
Only activation tensors are exchanged between rack nodes,
while model weights remain local to the node that owns the stage.

\noindent\textbf{Intra-Server Pipeline via WA Separation.}
Within each rack node, two CPU sockets are available (named CPU1 and CPU2).
We treat each CPU socket as a \emph{node} in the architectural sense
defined in \S\ref{sec:sys-paradigm}.
As discussed in \S\ref{sec:wa-decouple},
The execution model partitions each transformer layer into two sequential stages:
(i) weight-centric operators (QKV projection and FFN),
and (ii) attention-centric operators.
CPU1 is designated as the \emph{weight node},
and executes weight-centric operators, while
CPU2 is designated as the \emph{attention node},
and maintains KV-cache state while executing attention computation.

This intra-server inter-node pipeline prevents dynamically growing KV-cache
from competing with reusable model weights for cache capacity
within the same socket,
while retaining a low communication overhead.

\noindent\textbf{Tensor Parallelism within a Socket.}
Inside each socket-node, tensor parallelism (TP)
is realized across physical cores.
Each core is treated as a resource domain,
providing an execution context with private L1/L2 caches.
Matrix–vector multiplications are partitioned across cores,
where each core computes a disjoint shard of output channels.
This aligns the scheduling granularity of TP
with the smallest cache-resident compute domain.

\subsection{Cache-resident Computational Kernels}
\label{sec:impl_kernels}

LLM decoding is dominated by matrix--vector multiplication (GEMV) with
low data reuse in both linear projection and attention.
To minimize data movement across memory hierarchies, large structures
such as weights and KV cache are streamed from LLC, while KB-scale
activation embeddings are kept in L1.

\noindent\textbf{GEMV Kernel.}
During decoding, QKV projection and FFN layers reduce to a sequence of
independent inner products, so output channels are
partitioned across cores to perform GEMV on its assigned weight shard.

Since weight matrices have significantly larger
footprint than private caches,
we do not attempt to retain them in L1/L2.
For example, three $4096\times4096$ INT8 projection matrices require
48MB.
Each shard is 512KB when partitioned across 96 cores,
which comfortably fits within the 12MB LLC slice.

In contrast, the activation (embedding) vector is sufficiently small,
around 1$\sim$8KB depending on the model,
to reside entirely within the 32KB L1 data cache.
Under tensor parallelism, each core computes a disjoint weight shard
but consumes the same activation vector.
We therefore keep a per-core copy of the activation in L1,
so that all high-frequency reuse occurs at the highest-bandwidth level.

This organization establishes a simple design principle:
LLC is treated as a streaming layer for large-footprint data,
while L1 is reserved for high-frequency reuse.
Computation is structured so that data cross the LLC–core boundary
as few times as possible.

The same bandwidth consideration also explains the synchronization
design in the feed-forward (FFN) stage, as illustrated in
Fig.~\ref{fig:lock}, which exposes a fundamental trade-off between
synchronization overhead and memory traffic:
One option is to adopt an outer-product–style partial accumulation,
where each core computes independent partial updates and
avoids early synchronization.
While this reduces synchronization cost,
it would require repeatedly streaming weight tiles
or materializing intermediate vectors in LLC,
thereby increasing LLC-to-core traffic.
Given that LLC bandwidth and latency remain substantially
inferior to L1, repeated weight streaming quickly dominates
execution time.
In contrast, a bounded fan-in synchronization scheme
introduces moderate coordination overhead
but ensures that weight tiles are streamed only once,
preserving L1 reuse and minimizing LLC traffic.

In the cache-resident regime targeted by this design,
memory traffic across the LLC boundary is more expensive
than synchronization.
Therefore, we explicitly favor controlled synchronization
over additional LLC streaming.

To improve locality further, weight shards are partitioned and
initialized in a bank-aware manner so that each
core predominantly accesses
LLC slices that are physically close to it.
This placement is established during cache warmup
at system initialization---
when each core first touches its assigned weight shard,
that first touch determines the shard's home cache bank,
which is crucial for keeping it near the consuming core.
See \S\ref{sec:impl_runtime} for more details on
cache-aware thread placement and initialization.

\noindent\textbf{Attention Kernel.}
Attention computation is executed on the attention socket using
a Flash-style kernel~\cite{flashattention}.
KV-cache blocks are processed in a tiled fashion,
computing attention scores and value aggregation without materializing
large intermediate matrices.
Since KV-cache footprint grows with context length,
we similarly rely on LLC streaming for KV blocks while maintaining
query vectors in private cache.

\subsection{Customized Thread Pool Runtime}
\label{sec:impl_runtime}

All the cache-residency and scheduling mechanisms discussed above
require precise control over execution placement,
synchronization structure, and task ordering.
We therefore implement a static thread pool runtime
instead of relying on existing frameworks such as OpenMP~\cite{openmp}.

OpenMP provides general-purpose dynamic scheduling, task stealing,
and configurable affinity policies.
However, its synchronization model operates primarily at global-barrier
granularity.
Under tensor-parallel execution, these primitives introduce additional
overhead and do not map well to our sub-operator coordination scheme.
Prior microsecond-scale runtime systems show the same pattern:
once tasks become short-lived, generic scheduling overhead rapidly
becomes visible~\cite{zygos,arachne}.

In addition, OpenMP exposes only a limited set of coarse-grained
affinity policies (e.g., \texttt{spread} or \texttt{close}) rather
than deterministic per-core binding control.
Our design relies on fixed shard-to-core mapping and predictable cache
bank allocation, so this interface is not sufficient.

\noindent\textbf{Static Per-Socket Thread Pools.}
We instantiate one static thread pool per socket.
Threads are created once at initialization
and permanently pinned to physical cores
to repeatedly execute shard-local computation.
Our design benefits from this static thread pool in two ways:

First, model execution follows a fixed dependency structure
with deterministic shard-to-core mapping.
Since each core is dedicated to the same weight shard of
an operator,
task progression reduces to lightweight state transitions
within a fixed execution loop,
rather than general-purpose task scheduling.
Even a minimal task queue requires shared atomic counters
or lock-protected data structures to coordinate producers and consumers.
Such atomic operations introduce non-trivial
contention and cache-line bouncing across cores,
especially in this highly concurrent setting.
By reducing execution to state transitions
within a fixed execution loop,
we avoid most shared atomic data structures.

Second, because each thread is permanently bound to a specific core,
the same pinned thread can first touch the weight shard
that it later consumes during cache warmup.
As discussed above, this first touch follows the same locality
principle as first-touch NUMA placement~\cite{next-touch} and
helps keep the shard on LLC slices closest to the consuming core.

Two thread pools are isolated per socket.
No thread migrates across sockets,
and each socket-node maintains an independent execution context.
This per-socket isolation is essential for enforcing
the weight–attention (WA) separation:
the weight socket executes weight-centric operators,
while the attention socket maintains KV-cache state
and executes attention kernels,
without cross-socket interference at the thread level.

\noindent\textbf{Hierarchical Synchronization.}
To implement sub-operator asynchronicity,
we employ a hierarchical synchronization structure
built on lightweight atomic integers.
Instead of maintaining a single global atomic counter
that all cores update,
we partition synchronization across CCD boundaries.

Each CCD maintains a local atomic counter
shared only by cores within that CCD.
Threads first synchronize within their local CCD,
propagating completion signals to a per-CCD coordinator.
Only after intra-CCD synchronization completes
do representatives from each CCD participate
in a second-level synchronization step.
This follows the same basic principle as scalable shared-memory
synchronization and NUMA-aware locking~\cite{mcs-lock,lock-cohorting}:
keep highly contended state local and limit cross-domain ownership
transfer.

This two-level design explicitly bounds the fan-in degree
at each synchronization point.
In a flat global barrier, a single synchronization variable
is updated by all participating cores.
The fan-in degree in this case equals the total number of cores,
and each additional participant introduces another potential
owner of the corresponding cache line.
As the atomic counter is repeatedly modified,
its cache line must transfer ownership among all participants,
causing cache-line bouncing proportional to the fan-in degree.

In contrast, our hierarchical design limits fan-in locally.
Within each CCD, only cores in that CCD participate
in the first-level synchronization,
so cache-line ownership transfers remain confined
to that local domain.
At the second level, only one representative per CCD
participates in the cross-CCD synchronization,
reducing the number of bouncing parties to the CCD representatives
rather than all cores.

Bounding fan-in therefore directly bounds
the number of ownership transfers and coherence traffic,
instead of allowing them to scale with total core count.

There is a tradeoff between synchronization
overhead and memory traffic:
As discussed in the FFN implementation (Fig.~\ref{fig:lock}),
we choose global hierarchical synchronization,
which ensures that weight shards are streamed only once
and avoids execution strategies that would otherwise
increase repeated LLC traffic.
In the cache-resident regime targeted by this design,
bouncing shared cache lines is cheaper than
streaming weight tiles multiple times from the LLC.

\subsection{Inter-node Communication}

As discussed above, within a server node the weight and attention
sockets communicate through shared memory and lightweight
synchronization between two pinned thread pools.
The rest of this section therefore focuses on communication among
cluster nodes.

Our profiling shows that a naïve Ethernet connection requires roughly
$0.5$ms per hop to transfer a simple 4KB embedding packet.
This overhead is already comparable to the execution time of
cache-resident operator kernels.
As prior low-latency dataplane systems show~\cite{ix}, once the
compute path shrinks to microseconds, software overhead in the
communication stack becomes part of the critical path.
To reduce communication latency, the prototype adopts InfiniBand RDMA to
transfer activation tensors between consecutive pipeline stages.
Inter-node communication occurs only when embeddings produced by one
pipeline stage are forwarded to the next stage.

To minimize runtime overhead, RDMA queue pairs are established during
initialization and reused throughout execution.
Communication buffers are pre-registered so that transfers bypass the
kernel networking stack and avoid dynamic memory registration.
Since pipeline stage mappings are static, the communication topology
also remains fixed during decoding, eliminating connection setup and
routing overhead on the critical path.

Our evaluation shows that RDMA communication across four pipeline nodes
incurs about 50$\mu$s (roughly $0.5\%$ of the end-to-end latency) in
end-to-end LLaMA-3B decoding.
Refer to the execution-time breakdown in the evaluation section for
details.

\begin{table}[t]
    \centering
    \begin{tabular}{lrrrr}
        \toprule
        & {\footnotesize\shortstack[c]{LLaMA-3.2\\3B}} & {\footnotesize\shortstack[c]{LLaMA-2\\7B}} & {\footnotesize\shortstack[c]{Qwen-3\\8B}} & {\footnotesize\shortstack[c]{LLaMA-2\\70B}} \\
        \midrule
        \#Layers & 28 & 32 & 36 & 80 \\
        \#Sockets & 4+1 & 8+1 & 9+1 & 80+1 \\
        Layers/socket & 7 & 4 & 4 & 1 \\
        INT8 wt. (GB) & 3.21 & 6.74 & 8.19 & 68.98 \\
        \bottomrule
    \end{tabular}
    \vspace{0.1in}
    \caption{Model partitioning parameters. ``+1'' indicates an additional serving socket for embedding/argmax stages.}
    \label{tab:model-partition}
    \vspace{-0.3in}
\end{table}

\section{Experiment Setup} \label{sec:method}

This section describes the experimental platform, models,
baselines, and evaluation metrics used in our study.

\noindent\textbf{Hardware Platform.}
We prototype the design on a 4-rack-node server cluster with
AMD EPYC 9684X processors. Each server node contains
two CPU sockets with built-in motherboard interconnection.
The aggregated LLC capacity
per socket is 1152MB (96 cores $\times$ 12MB),
providing sufficient
capacity for cache-resident execution.

Each node is equipped with an NVIDIA ConnectX-7 NIC connected through PCIe; nodes communicate via a non-blocking Ethernet switch at 400 Gbps.
Each node also has 1.5TB DDR5 memory
($\approx$ 400GB/s memory bandwidth).

\noindent\textbf{Models and Quantization.}
The prototype is evaluated on three representative models, LLaMA-3.2-3B,
LLaMA-2-7B, and Qwen-3-8B, plus a larger
LLaMA-2-70B configuration for performance simulation.

All models are evaluated in a fully INT8 configuration, including the
KV cache. Prior work such as SmoothQuant~\cite{smoothquant} shows that
INT8 LLM inference can preserve accuracy sufficiently for practical
deployment.
This setting enables x86 VNNI acceleration for dot-product-heavy kernels.
Table~\ref{tab:model-partition} summarizes each pipeline partition.
Unless otherwise stated, transformer layers are evenly split across pipeline
sockets. INT8 indicates one byte per parameter. Socket counts are
written as ``compute + serving'', where an extra serving socket
is needed for input embedding and output argmax,
while ``Layers/socket'' counts only the transformer
layers assigned to each compute socket.

\noindent\textbf{Baselines.}
We compare the prototype with \textbf{llama.cpp}\cite{llama-cpp}, a widely used
LLM inference engine with strong CPU support, which follows a
conventional operator-centric execution model and provides
a CPU backend for inference.

To ensure a fair comparison, both systems are evaluated on
the same hardware platform using identical model checkpoints
and quantization configurations.

\section{Evaluation} \label{sec:eval}

We use the hardware platform described in \S\ref{sec:method}.
Unless otherwise stated, we compare against \texttt{llama.cpp} running on an
equally provisioned cluster with the same models, quantization configuration,
context lengths (from $1024$ to $4096$), and batch sizes (from $1$ to $32$).
With regard to correctness, both the Llama-3.2-3B and Llama-2-7B prototype
deployments exactly match the corresponding original model outputs.

Our evaluation proceeds in three steps.
The default configuration uses single-socket execution
(i.e., no weight-attention separation) and the specialized thread pool.
We first measure end-to-end performance on the available Llama-3.2-3B and
Llama-2-7B deployments.
We then use those measurements to validate an analytical model that captures
the dominant performance trends of the prototype from per-node profiles.
Finally, we use that validated model together with controlled single-node
ablations to study scaling behavior and isolate the contributions of the
specialized thread pool and weight-attention (WA) separation.
This staged methodology separates direct end-to-end measurement from
model-based extrapolation and mechanism-focused ablation, while keeping all
three tied to the cache-pressure bottleneck identified in Challenge~1 in
\S\ref{sec:scalability}.

The key findings are as follows:
\begin{itemize}
    \item \textbf{Measured end-to-end.} On the Llama-3.2-3B and
    Llama-2-7B deployments at context length $4096$, the prototype reduces
    end-to-end TPOT by $2.04\times$--$11.51\times$ over
    \texttt{llama.cpp}, with the largest gains at small batch sizes.
    \item \textbf{Validated trend model.} Our analytical model captures the
    batch-scaling trend of measured end-to-end TPOT. For the Llama-3.2-3B
    deployment, measured-to-estimated TPOT stays close to 1
    ($1.26\times$--$1.48\times$), while for Llama-2-7B it is
    systematically higher ($1.15\times$--$1.52\times$), indicating that
    the model slightly
    underestimates TPOT on the deeper pipeline while preserving the correct
    trend despite additional inter-node interference and cache thrashing.
    \item \textbf{Thread pool ablation.} The specialized static thread
    pool usually removes a mostly fixed runtime overhead, saving tens of
    microseconds per transformer block. Its geomean speedup is modest
    ($1.05\times$--$1.16\times$ across models) but reaches up to
    $1.56\times$ at small batches.
    \item \textbf{Weight-Attention separation ablation.} WA separation is
    workload dependent. At context length $4096$, the dual-socket design is
    nearly neutral on Llama-3.2-3B ($1.00\times$ geomean) and provides
    moderate gains on the larger Llama models ($1.13\times$ on Llama-2-7B
    and $1.16\times$ on Llama-2-70B). Figure~\ref{fig:wa-decouple-perf}
    suggests why: separating weight-centric operators from attention prevents
    the two phases from polluting the same socket-level cache, so extra
    sockets can be translated into lower latency.
\end{itemize}

\subsection{End-to-end Performance}

End-to-end experiments provide the ground-truth reference for our design.
Due to limited hardware availability, we report them only for the
representative Llama-3.2-3B and Llama-2-7B deployments.
We focus on TPOT decoding because it occupies most of the
inference time.
The current prototype prioritizes decoding performance, and we leave a
detailed treatment of prefill to future work.

\begin{table}[t]
    \centering
    \small
    \hspace{-0.25in}
    \begin{tabular}{lrrrrrr}
        \toprule
        Batch size & 1 & 2 & 4 & 8 & 16 & 32 \\
        \midrule
        \multicolumn{7}{@{}l}{\textbf{(a) TPOT (ms) - LLaMA-3.2-3B}} \\
        \texttt{llama.cpp} & 48.6 & 49.0 & 53.7 & 82.1 & 138.5 & 215.8 \\
        Measured & 4.2 & 8.4 & 15.7 & 24.4 & 43.8 & 76.3 \\
        Speedup (x) & 11.51 & 5.83 & 3.41 & 3.37 & 3.16 & 2.83 \\
        Estimated & 2.9 & 5.8 & 11.9 & 19.3 & 32.7 & 59.8 \\
        Meas./Est. (x) & 1.48 & 1.46 & 1.32 & 1.26 & 1.34 & 1.28 \\
        \midrule
        \multicolumn{7}{@{}l}{\textbf{(b) TPOT (ms) - LLaMA-2-7B}} \\
        \texttt{llama.cpp} & 82.5 & 82.6 & 111.8 & 146.1 & 227.4 & 378.7 \\
        Measured & 7.9 & 17.8 & 29.7 & 63.2 & 87.6 & 185.8 \\
        Speedup (x) & 10.43 & 4.64 & 3.77 & 2.31 & 2.60 & 2.04 \\
        Estimated & 6.9 & 14.9 & 24.2 & 43.0 & 69.0 & 122.0 \\
        Meas./Est. (x) & 1.15 & 1.20 & 1.23 & 1.47 & 1.27 & 1.52 \\
        \bottomrule
    \end{tabular}
    \vspace{0.1in}
    \caption{End-to-end performance of LLaMA-3.2-3B and LLaMA-2-7B at context length $4096$.}
    \label{tab:e2e-summary}
    \vspace{-0.1in}
\end{table}

Table~\ref{tab:e2e-summary} summarizes the context-length-$4096$ end-to-end
TPOT results.
For Llama-3.2-3B, the TPOT speedup ranges from $11.51\times$ at batch size $1$
to $2.83\times$ at batch size $32$; for Llama-2-7B, it ranges from
$10.43\times$ to $2.04\times$ (Table~\ref{tab:e2e-summary}(a) and (b)).
In both models, the relative advantage is strongest at small batch sizes and
shrinks as batch size grows, which is consistent with the expectation that
larger batches better amortize the cost of loading weights from memory and
therefore narrow the gap to the prototype.

\subsection{Analytical Model Validation}

Because our available hardware only supports end-to-end deployment up to the
Llama-2-7B configuration, we next validate a simple analytical model before
extrapolating to larger models.
Our goal is not exact point prediction, but faithful trend prediction under the
same KV-cache pressure as the deployed system, so that profiled single-layer
latency remains representative of the actual decoding regime.

\noindent\textbf{The number of in-flight instances.} As shown in
Table~\ref{tab:model-partition}, each model requires $n+1$ sockets to support
end-to-end decoding, where $n$ is the number of transformer-compute sockets and
$+1$ accounts for the dedicated embedding/argmax socket.
However, the load across these sockets is highly asymmetric: our profiling
shows that embedding/argmax typically takes $\sim\!10\mu$s, whereas a
transformer socket typically takes hundreds of microseconds to a few
milliseconds.
We therefore model the system with only $n$ in-flight instances rather than
$n+1$.
This choice may leave a small amount of throughput on the table, depending on
the share of embedding/argmax latency in TPOT, but it better matches the
latency-oriented scheduling policy of the current prototype.
Accordingly, \#Sockets below refers only to transformer sockets (without the
extra embedding/argmax socket).

We estimate the model with the following decomposition:
\[
\begin{array}{@{}l@{}}
\text{Throughput} = \displaystyle\frac{\text{(Batch size)}}{\text{(Per-stage latency)}}, \\
\text{TPOT} = \#\text{Sockets} \times \text{((Per-stage latency) + (Nw. latency))} \\
\phantom{\text{TPOT} =}\ + \text{Embedding}.
\end{array}
\]

Here, the per-stage latency is the measured execution time of one socket
processing all stages assigned to it in Table~\ref{tab:model-partition}, and
the network latency is measured as $5\mu$s per communication.
Because decoding traverses the transformer pipeline together with the dedicated
embedding/argmax stage, we model \#Sockets communications per decoded token.

The TPOT blocks in Table~\ref{tab:e2e-summary}(a) and
Table~\ref{tab:e2e-summary}(b) show that this model captures the measured
batch-scaling trend, but with some residual error.
Across the representative deployments, the Llama-3.2-3B
measured-to-estimated ratios remain close to 1
($1.26\times$--$1.48\times$), whereas the Llama-2-7B ratios are
systematically higher ($1.15\times$--$1.52\times$), meaning that the model
slightly underestimates TPOT on the deeper pipeline.
We attribute this stable Llama-2-7B gap to additional inter-node
interference and cache thrashing in the full end-to-end system, which are not
captured by the simplified layer-plus-network decomposition.
Even so, the model preserves the correct batch-scaling trend and order of
magnitude.
We therefore use it as a trend-level extrapolator, rather than an exact
absolute predictor, for the larger-model results below.

With this validation in place, the remainder of the evaluation focuses on
controlled single-node studies, which let us vary cache pressure,
working-set composition, and placement decisions while remaining faithful to
the underlying performance bottleneck.

\begin{figure}[t]
    \centering
    \hspace{-0.25in}
    \includegraphics[width=1.05\linewidth]{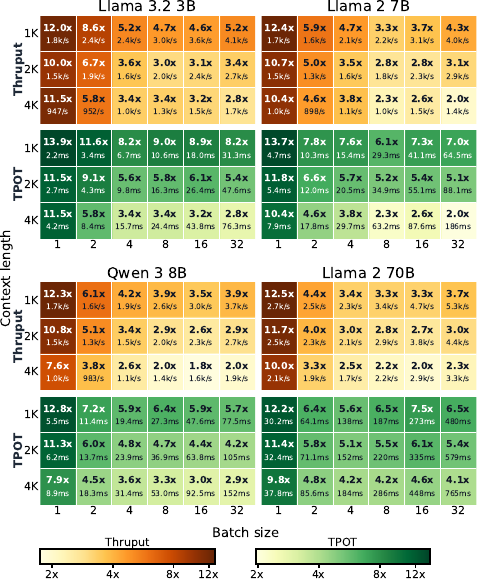}
    \Description{Heatmap comparing prototype and llama.cpp performance across batch sizes and context lengths.}
    \vspace{-0.12in}
    \caption{Model-based throughput and TPOT speedups over \texttt{llama.cpp}
    across context lengths and batch sizes for four models. The speedup is on
    top of each cell, and the corresponding absolute throughput (tokens/s) or
    TPOT (ms) is at the bottom.}
    \label{fig:tpot-and-thruput}
    \vspace{-0.1in}
\end{figure}

\begin{figure*}[t]
    \centering
    \includegraphics[width=0.9\textwidth]{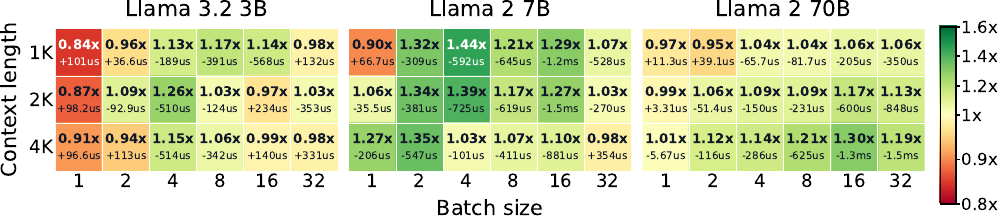}
    \Description{Heatmap showing single-socket versus dual-socket performance comparison.}
    \vspace{-0.1in}
    \caption{Effect of weight-attention separation on per-transformer-block
    latency in Llama-family models. The speedup is on top of each cell, and
    the absolute time saving per transformer block ($\mu$s) is at the bottom.}
    \label{fig:wa-decouple-perf}
    \vspace{-0.1in}
\end{figure*}

\begin{figure}[t]
    \centering
    \hspace{-0.25in}
    \includegraphics[width=1.05\linewidth]{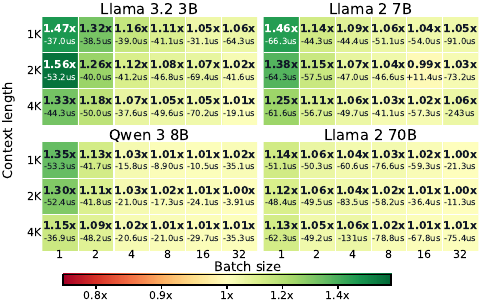}
    \Description{Heatmap comparing OpenMP versus specialized runtime performance.}
    \vspace{-0.12in}
    \caption{Customized thread pool versus OpenMP in a single-socket,
    per-transformer-block comparison. The speedup is on top of each cell, and
    the absolute time saving per transformer block ($\mu$s) is at the bottom.}
    \label{fig:omp-vs-pool}
    \vspace{-0.1in}
\end{figure}

\subsection{Sensitivity to Context Length and Batch Size}

After validating the analytical model, we can directly
use the per-layer latency profiles to extrapolate the
end-to-end performance of larger models and more configurations.
Figure~\ref{fig:tpot-and-thruput} shows that the prototype achieves up
to $13.9\times$ TPOT speedup and $12.5\times$ throughput speedup over \texttt{llama.cpp}
under this extrapolation.
Across the full grid, the geomean speedup remains substantial at
$3.7\times$--$5.0\times$ for throughput and $5.3\times$--$6.7\times$ for TPOT,
depending on the model.

The relative speedup shrinks as batch size increases, which is
consistent with the measured end-to-end trend in
Table~\ref{tab:e2e-summary}: larger batches better amortize weight-loading
overhead in the baseline and therefore narrow the relative advantage of the prototype.
Further, the per-stage execution time often increases sharply from batch size 1
to 2, then grows more gently from batch size 4 onward.
This suggests that batch size 2 already exposes enough data parallelism to keep
the 96 AVX-512 cores well utilized, while larger batches increasingly help
amortize weight movement from LLC rather than scaling compute time
proportionally.

Large batches can improve hardware utilization, but they require request
accumulation and often worsen queueing and tail latency in user-facing serving
systems.
This makes the small-batch regime (1--8) the practically relevant operating
point for latency-sensitive serving, and Figure~\ref{fig:tpot-and-thruput}
shows that this is also where the proposed design is strongest.

\subsection{Effect of the Specialized Thread Pool}

We compare our specialized static thread pool against an OpenMP-based runtime
in the controlled single-socket study of Figure~\ref{fig:omp-vs-pool}.
The specialized runtime usually removes a mostly fixed latency overhead per
transformer block, typically tens of microseconds and occasionally more than
100$\mu$s.
As a result, its overall geomean speedup is modest ($1.05\times$--$1.16\times$
across models), but it reaches up to $1.56\times$ when batches are small.
In that regime ($1$--$8$), a transformer block itself takes only hundreds of
microseconds, so removing tens of microseconds translates into a large relative
improvement.
As batch size increases, however, block latency moves into the millisecond
range, so the same fixed microsecond-scale saving is amortized and shrinks to
only a few percent at the largest batches.

This is consistent with our discussion in \S\ref{sec:impl_runtime}.
Our runtime uses permanently pinned threads with deterministic
shard-to-core mapping to avoid fixed runtime overheads from a generic
runtime, such as affinity setup and global barrier management.
Once cache residency shortens kernel service time, they become visible in the
small-batch regime; once computation dominates at larger batches, the same
absolute saving remains but its relative impact becomes much smaller.

\subsection{Effect of Weight-Attention Separation}

We next isolate the effect of WA separation by comparing the full design,
which places weight-centric operators and attention on different sockets,
against a colocated variant that executes both on the same socket.

\begin{figure}[b]
    \centering
    \makebox[\linewidth][r]{%
        \includegraphics[width=0.92\linewidth]{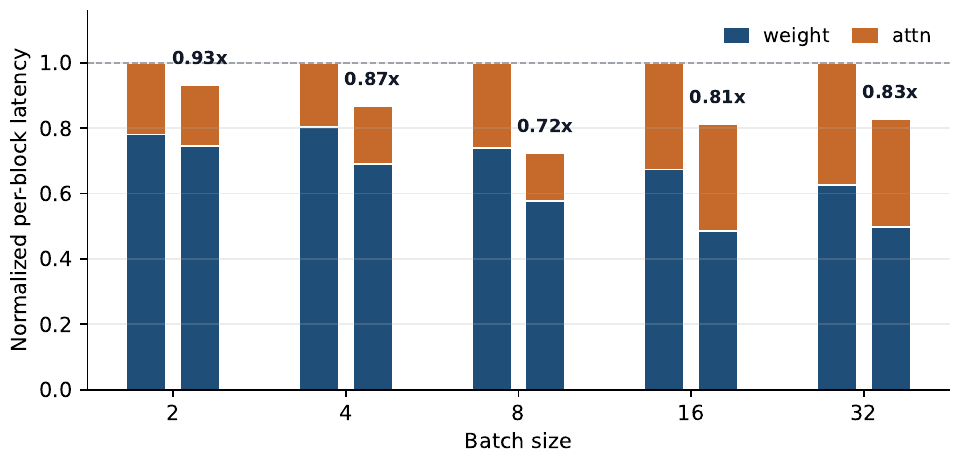}
    \Description{Bar chart showing weight-attention execution breakdown for Llama-2-70B model.}
    }
    \vspace{-0.08in}
    \caption{Llama-2-70B at ctx=4096 weight attention separation breakdown.
    The left bars are single-socket (=1), and the right bars are
    weight-attention separated.}
    \label{fig:wa-decouple-breakdown}
    \vspace{-0.08in}
\end{figure}

Figure~\ref{fig:wa-decouple-perf} shows that the benefit depends strongly on
cache pressure, which is jointly determined by model size, context length, and
batch size.
At context length $4096$, the geomean per-block speedup is essentially neutral
for Llama-3.2-3B ($1.00\times$), rises to $1.13\times$ for Llama-2-7B, and
to $1.16\times$ for Llama-2-70B.
When both batch size and context length are small, the combined weight and
KV-cache working set is still modest enough to fit in LLC even without
separation.
The coordination overhead of the dual-socket design can then outweigh the
cache benefit, and Llama-3.2-3B indeed shows a slowdown at many points.
As the working set grows, however, the benefit becomes more systematic.
Llama-2-7B shows a clear but moderate improvement, while Llama-2-70B becomes
more consistently positive in the high-pressure region.
The pattern suggests that weight-centric operators and attention interfere much
less once they stop sharing the same socket-level cache and memory path.
In the colocated design, their footprints compete in the same cache hierarchy
and can evict one another; with WA separation, the two phases no longer
mutually pollute the same socket, so the larger working set can be sustained
with less interference.
Figure~\ref{fig:wa-decouple-breakdown} zooms into the Llama-2-70B case
at context length 4096. Both attention and weight-centric
operators are accelerated because the growing KV cache no longer evicts
weight-resident data, and attention no longer competes for cache
resources as in the single-socket design.

This latency improvement is therefore not free: it opens a tradeoff between
per-request latency and throughput efficiency at a fixed socket budget.
WA separation dedicates one socket to weight-centric computation and another to
attention, which doubles the socket budget for a decoding stream.
Because the resulting latency speedup is usually sublinear in socket count, the
throughput efficiency per socket is reduced even when latency improves.
We therefore view WA separation as a good way to scale latency with additional
resources: once cache pressure is high enough, spending another socket to keep
weights and attention from interfering is often worthwhile even if the
throughput gain per socket is sublinear.
The larger-model results are therefore best read as support for the underlying
cache-pressure hypothesis, but with moderate and workload-dependent gains
rather than a universal large win.

\section{Discussion}

\subsection{Related Works}

\noindent\textbf{Dedicated Resource Allocations:}
Prior systems have explored phase-aware LLM serving~\cite{dist-serve,splitwise,sarathi-serve},
separating prefill and decoding or scheduling them differently to better match
distinct compute and memory characteristics on GPUs.
In contrast, this work approaches this issue through
weight--attention separation to specialize for the different compute
and memory patterns.

\noindent\textbf{KV-cache Management:}
Recent systems optimize inference by managing the growing KV cache more
efficiently. PagedAttention\cite{pagedattention} reduces
fragmentation through a paged KV layout, vAttention~\cite{vattention}
retains a contiguous virtual layout while allocating physical memory
on demand, and InfiniGen~\cite{infinigen} reduces long-context fetch
overhead by selectively loading KV entries.
This work makes a different CPU-oriented tradeoff:
because extra instructions and pointer indirections fall directly on
our critical path, we cannot adopt a PagedAttention-style
address-translation layer in the attention loop.
Instead, we keep the KV layout simple and isolate KV interference from
weight-resident execution through weight--attention separation.

\noindent\textbf{Cache-resident Inference:}
The proposed design shares a similar SRAM-resident intuition with specialized accelerators
such as Groq~\cite{groq}, but operates under a fundamentally different set of
constraints.
Recent work such as WaferLLM~\cite{waferllm} also redesigns inference
around massive on-chip memory, but on wafer-scale accelerators rather
than commodity CPUs.
Groq relies on a highly specialized LPU architecture and a tightly integrated
software stack for large-model execution, which allows it to push tensor
parallelism and inter-device communication optimization much further.
Systems such as FlexGen~\cite{flexgen} instead target limited-resource
inference through offloading across GPU, CPU, and disk.
In contrast, this work is built entirely within the commodity CPU ecosystem, where
cache coherence, fixed hardware topology, and general-purpose runtime overhead
impose much stronger constraints.
Our contribution is therefore not to replicate accelerator-style execution,
but to show that a related locality-centered principle can be realized in
software on widely available CPUs.

\noindent\textbf{CPU Inference:}
Before the LLM era, CPUs were widely used for inference~\cite{fb-datacenter-hpca-2018},
but the scale and performance requirements of LLMs have led to a shift
towards GPU-only serving for dense and conventional models.
T-Mac enables CPU-based LLM inference through a table-lookup approach~\cite{tmac},
but is limited to small models and short contexts due to its reliance on a large
precomputed table of activations. The proposed design, in contrast, is designed to scale to
larger models and longer contexts by optimizing execution on commodity CPUs.

\subsection{Future Works}

Several directions could extend this design. Continuous batching~\cite{orca,sarathi-serve}
would require online KV-cache reclamation and batch reconfiguration while
preserving socket-local hot state.
Mixture-of-Experts models would introduce routing-dependent communication and
would require topology-aware expert placement to keep sparse activation from
turning into cross-socket traffic.
Another direction is extreme latency optimization for interactive workloads
such as AI-assisted coding, prioritizing per-request token latency through
dedicated cores and lower-overhead synchronization.

\subsection{Conclusion}

This work rethinks LLM serving as both a systems problem and a resource-allocation
problem. For an important class of workloads, it shows that commodity CPUs can
sustain serving when execution is organized around cache residency and
topology-aware coordination, rather than assuming GPUs as the default platform.
This can reserve scarce accelerator capacity for training and fast iteration
while using CPUs for deployment. Although the design is instantiated on CPUs, it also
offers scalability guidance for cache-resident accelerators, where bounded
on-chip memory makes resource isolation, locality preservation, and
topology-aware coordination equally central.

\bibliographystyle{plain}
\bibliography{ref}

\end{document}